\documentclass{article}
\usepackage{authblk}

\begin{document}

\title{It Will Never Work in Theory}

\author[1]{Greg Wilson}
\author[2]{Jorge Aranda}
\author[3]{Michael Hoye}
\author[4]{Brittany Johnson}

\affil[1]{Third Bit, Toronto, ON, Canada}
\affil[2]{Workday, Victoria, BC, Canada}
\affil[3]{Pluralsight, Toronto, ON, Canada}
\affil[4]{George Mason University, Fairfax, VA, USA}

\date{February 2024}

\maketitle

\begin{abstract}
  \noindent
  We have been trying to get software engineering researchers and
  practitioners to talk to one another for over a decade.  This paper
  describes what we have done, assesses our impact, and recommends an
  approach that we hope will have greater success.
\end{abstract}

\section{History}

In 2011, Jorge Aranda (then a graduate student at the University of
Toronto) and Greg Wilson (who had recently left academia for the third
time) began posting short reviews of software engineering research
papers to \emph{It Will Never Work in Theory} \cite{NWIT}.  Both were
frustrated by the gulf between what researchers studied and what
practitioners in industry cared about, and conversely by how little
attention working programmers paid to results that \emph{were}
relevant and useful.

Aranda, Wilson, and a handful of other contributors posted almost 90
reviews over the next three years (Table~\ref{postings}).  They
selected papers that they believed practitioners were most likely to
find interesting and/or actionable, that were not hidden behind the
Great Paywall of Academia, and that had a strong bias toward empirical
studies (both quantitative and qualitative).

A handful of posts attracted half a dozen comments or a passing
mention on social media, but neither of the communities the site hoped
to reach engaged in any significant way.  Posting continued
sporadically in 2014--15, and there was another burst of activity in
2016, but the project officially went on hiatus in December of that
year.  As the announcement at the time said, ``{\ldots}in the wake of
recent political events, our community's energy and attention should
be focused on more important things.''

\begin{table}
\begin{tabular}{rr}
\textbf{Year} & \textbf{Posts} \\
2011 &    43 \\
2012 &    30 \\
2013 &    16 \\
2014 &     4 \\
2015 &     3 \\
2016 &    33 \\
2021 &    96 \\
2022 &    74 \\
2023 &    82 \\
\end{tabular}
\caption{Posts per Year}
\label{postings}
\end{table}

Wilson revived the project in 2021, again focusing on open access
papers that practitioners were most likely to find actionable.  These
reviews were shorter---sometimes only a paragraph---and led up to the
first of three live events (Table~\ref{speakers}).  In collaboration
with Michael Hoye (then at Mozilla) and Prof.\ Brittany Johnson
(George Mason University), \emph{It Will Never Work in Theory} hosted
two sets of online lightning talks and one set co-located with the
Strange Loop conference \cite{SL}.  Speakers were given some coaching
on the differences between speaking to academic and non-academic
audiences, and then had 10 minutes to explain a research finding of
interest to practitioners.  Recordings of their talks were posted
online with transcripts in English and Spanish (the latter created by
Yanina Bellini Saibene and her colleagues).

\begin{table}
\begin{tabular}{llr}
\textbf{Date} & \textbf{Format} & \textbf{Speakers} \\
April 2022 & online    &       22 \\
Sept 2022 & in person &        8 \\
April 2023 & online    &       22 \\
\end{tabular}
\caption{Speakers per Event}
\label{speakers}
\end{table}

Readership on the site hovered around 50--100 visits per post
(compared to 700--1000 visits \emph{per day} for another of the first
author's projects that hasn't been updated in over a decade
\cite{AOSA}).  Each of the lightning talk videos was viewed by
150--1000 people in the first 30 days after publication, but only a
handful ever reached out to the presenters.

\section{Retrospective}

This attempt to get researchers and practitioners to talk to each other has
been personally rewarding, but has had no impact on software engineering's
two solitudes.

Most software developers in industry have never heard of any findings
more recent than \cite{Brooks1975} (which few of them ever actually
read), and routinely dismiss studies as ``not statistically
significant'', even when those studies are carefully done and directly
relevant to their work. When awareness of research does seep into
developers' conversations it is usually as ammunition to support
preconceived notions rather than any sincere attempt to improve their
knowledge or practice of the art. For example, people who prefer
strongly-typed languages will broadcast the fact that some recent
paper has proven they're better without examining what the paper's
authors actually mean by ``proven'' and ``better''.

Likewise, those researchers whose papers we reviewed and who presented
at our lightning talks have been no more likely to attend non-academic
conferences than they were before. Research has had some impact on
developer tooling---for example, all of today's IDEs draw on work in
static analysis---but uptake has primarily been by individuals and in
their individual work, rather than at the larger scale of standardized
or commonly understood practices that other fields would call
``engineering''.

\section{Recommendations}

Twelve years after \emph{It Will Never Work in Theory} launched, the
real challenge in software engineering research is not what to do
about ChatGPT or whatever else Silicon Valley is gushing about at the
moment. Rather, it is how to get researchers to focus on problems that
practitioners care about and practitioners to pay attention to what
researchers discover.  This was true when we started, it was true ten
years ago \cite{Begel2014}, and it remains true today.

We believe the best time and place to bridge this divide is when we
have the attention of future researchers and practitioners, i.e., in
undergraduate programs. After all, if students leave academia without
having been exposed to both research methods and useful discoveries,
why would those who leave look to researchers later for help or
answers?

Software engineering faculty could, if they wanted, replace the team
programming project course that most students do in their third or
fourth year with one in which they design a small study or experiment,
collect data, analyze it, and figure out what (if anything) they've
proven. Such a class would not disrupt other curricula, would give
students a chance to learn some practical data science, and would help
prepare them for graduate school (which is in professors' own
interests as well). Crucially, students would be more likely to
understand and value researchers' findings having done a little
research of their own.

\section{Conclusion}

The comedian W.C.~Fields once said,
``If at first you don't succeed, try, try again.
Then quit.
There's no point in being a damn fool about it.''
Thirteen years after our first post,
it is clear that
our attempts to bridge the gulf between research and practice haven't worked.
We look forward to hearing what actionable plans others have
that will find real support from both communities.

\section{Acknowledgments}

We would like to thank the universities, companies, and granting
agencies that supported this work, but unfortunately none did.  We
are, however, very grateful to everyone who contributed to this
project over the years.

\section*{About the Authors}

\noindent
\textbf{Greg Wilson} is a senior software engineering manager at Deep Genomics.
Contact him at gvwilson@third-bit.com.

\vspace{\baselineskip}

\noindent
\textbf{Jorge Aranda} is a principal software engineer at Workday.
Contact him at jorge.aranda@cuevano.ca.

\vspace{\baselineskip}

\noindent
\textbf{Michael Hoye} is the Director of Developer Engagement at Pluralsight by day,
a systems adminstrator by birth,
and a community manager and entrepreneur out of habit.
Contact him at mike-hoye@pluralsight.com.

\vspace{\baselineskip}

\noindent
\textbf{Brittany Johnson} is an assistant professor
in the Department of Computer Science at George Mason University
where she directs the Interdisciplinary Software Practice Improvement Research and Development (INSPIRED) Lab.
Contact her at johnsonb@gmu.edu.


\begin{thebibliography}{1}

\bibitem{NWIT}
``{It Will Never Work in Theory},'' https://neverworkintheory.org/, accessed
  January 2024.

\bibitem{SL}
``{Strange Loop},'' https://thestrangeloop.com/, accessed January 2024.

\bibitem{AOSA}
``{The Architecture of Open Source Applications},'' https://aosabook.org/,
  accessed January 2024.

\bibitem{Brooks1975}
F.~P. {Brooks Jr.}, \emph{The Mythical Man-Month: Essays on Software
  Engineering}.\hskip 1em plus 0.5em minus 0.4em\relax Addison-Wesley, 1975.

\bibitem{Begel2014}
A.~Begel and T.~Zimmermann, ``Analyze this! 145 questions for data scientists
  in software engineering,'' in \emph{{ICSE'14}}, 2014.

\end{thebibliography}
\end{document}